# Integrative Deep Learning Framework for Parkinson's Disease Early Detection using Gait Cycle Data Measured by Wearable Sensors: A CNN-GRU-GNN Approach

Alireza Rashnu, Armin Salimi-Badr*

*Faculty of Computer Science and Engineering, Shahid Beheshti University, Tehran, Iran*

## ABSTRACT

Efficient early diagnosis is paramount in addressing the complexities of Parkinson's disease because timely intervention can substantially mitigate symptom progression and improve patient outcomes. In this paper, we present a pioneering deep learning architecture tailored for the binary classification of subjects, utilizing gait cycle datasets to facilitate early detection of Parkinson's disease. Our model harnesses the power of 1D-Convolutional Neural Networks (CNN), Gated Recurrent Units (GRU), and Graph Neural Network (GNN) layers, synergistically capturing temporal dynamics and spatial relationships within the data. In this work, 16 wearable sensors located at the end of subjects' shoes for measuring the vertical Ground Reaction Force (vGRF) are considered as the vertices of a graph, their adjacencies are modeled as edges of this graph, and finally, the measured data of each sensor is considered as the feature vector of its corresponding vertex. Therefore, The GNN layers can extract the relations among these sensors by learning proper representations. Regarding the dynamic nature of these measurements, GRU and CNN are used to analyze them spatially and temporally and map them to an embedding space. Remarkably, our proposed model achieves exceptional performance metrics, boasting accuracy, precision, recall, and F1 score values of 99.51%, 99.57%, 99.71%, and 99.64%, respectively.

## 1. Introduction

Parkinson's disease is a progressive neurological disorder characterized by the gradual degeneration of nerve cells in the brain, particularly those responsible for producing dopamine, a crucial chemical messenger involved in controlling movement [1]. It affects approximately 1% of the global population over the age of 60, making it one of the most prevalent neurodegenerative disorders worldwide. While the exact cause remains elusive, a combination of genetic predisposition, environmental factors, and age-related changes are believed to contribute to its development. The hallmark symptoms of Parkinson's disease include tremors, stiffness, slowed movements, and impaired balance, which can significantly impact an individual's quality of life and independence [2]. In conclusion, Parkinson's disease poses a significant burden on individuals and healthcare systems globally, highlighting the importance of early diagnosis and intervention. By recognizing the symptoms early on and initiating appropriate management strategies, individuals with Parkinson's disease can mitigate the progression of the condition and maintain a higher level of independence and quality of life.

Some studies, such as [3-6], have used physiological signals such as handwriting, tremors, gait cycle, and speech to diagnose Parkinson's disease. The gait cycle exhibits notable traits such as regularity, predictable patterns, and spatial-temporal attributes. Furthermore, compared to methodologies reliant on speech or handwriting, which predominantly capture non-motor symptoms, gait analysis allows clinicians to evaluate motor symptoms. This enables Visualization of the severity of subjects' impairment and offers a comprehensive assessment of their condition [7]. In addition, diagnosis using gait cycle signals is less expensive than methods such as [8, 9] that use complex genetic tests and brain imaging methods.

While there is no single definitive test for Parkinson's disease, currently, in the clinical setting, the diagnosis and severity rating of PD is based on the visual observation and UPDRS score prepared by [10]. Due to the subjectivity inherent in visual observation, assessments may be prone to bias. Consequently, there has been a recent shift in approach towards the adoption of machine learning (ML) algorithms. These algorithms aim to unveil concealed patterns within physiological signals, offering potential support to clinicians in their routine Parkinson's disease (PD) diagnoses. This not only enhances the precision of predictions but also streamlines the diagnostic process, reducing the time required for assessment.

Using non-optimal methods, such as feature extraction from different frequency bands, is necessary to train classical machine learning methods based on gait cycle signals. Therefore, this paper presents a deep learning-based approach for automatic feature extraction without any special knowledge with the aid of layers of convolutional neural network (CNN), graph neural network (GNN), and gait recurrent unit (GRU), which is called CGG. These layers allow

---

* Corresponding author. Tel.: +98-21-29904192

E-mail addresses: a.rashnu@mail.sbu.ac.ir (A. Rashnu), a_salimibadr@sbu.ac.ir (A. Salimi-Badr)

us to extract special, temporal, and complex dependency features. Finally, the acquired vector is inputted into dense layers to derive more condensed and advanced features by employing various nonlinear transformations. Our experiments illustrate that the proposed model outperforms the baselines. The main contributions of this article to the literature are as follows:

1. Paying attention to the complex features and the effect of each sensor position on its neighboring counterparts to improve the embedding process using the GNN layer;

2. Identifying effective sensors in the gait cycle of subjects with Parkinson's disease to improve and accelerate the patient's gait cycle pattern analysis based on the severity of the disease by assigning the attention coefficients to different sensors;

3. Having fewer parameters than other advanced baselines based on deep learning.

The rest of this paper is organized as follows: Section 2 reviews the state-of-the-art diagnosis of Parkinson's disease-based gait cycle dataset. In Section 3, the data description is presented. Section 4 illustrates our proposed method. Section 5 shows the experimental evaluations. Finally, section 6 concludes and explains the future works.

## 2. Related works

In recent years, the intersection of deep learning methodologies and gait analysis for Parkinson's disease diagnosis has emerged as a promising avenue for more accurate and timely detection. As researchers delve deeper into the complexities of gait patterns and their correlation with Parkinsonian symptoms, a plethora of studies have surfaced, showcasing diverse approaches ranging from traditional machine learning techniques to sophisticated neural network architectures.

In general, binary classification methods for distinguishing Parkinson's patients from healthy people based on vGRF time series can be divided into two categories: classical machine learning methods and deep learning-based methods. In classic machine learning models, a feature vector is extracted from the statistical analysis of each time series as an input to the model. In [4], four classification algorithms such as support vector machine (SVM), random forest (RF), decision tree (DT), and k-nearest neighbor (K-NN) have trained by extracting a set of features such as standard deviation, oscillation time, average value of the center of pressure, and step time variation coefficient from PD gait signals. According to their experiments, SVM with cubic Kernel performs better than other models. The process of classifying subjects into healthy or Parkinsonian classes in [11] includes four stages of data preprocessing, feature extraction, feature selection with the aid of a wrapper approach established using the RF algorithm, and then classification by using both supervised classification methods, including K-NN, RF, SVM, Naïve Bayes (NB), and unsupervised classification methods such as K-means and the Gaussian mixture model (GMM). In [12], sixteen-time domain features and seven frequency domain features have been extracted from GRF signals to train a hybrid model called Locally Weighted Random Forest (LWRF) for regression analysis. In [13], an interval type-2 neuro-fuzzy system has been proposed to extract fuzzy rules based on human-understandable features derived from the vGRF time series.

While effective in many applications, classical machine learning models present several demerits when applied to diagnosing Parkinson's disease using gait cycle data. Firstly, these models often require extensive feature engineering, where domain expertise is necessary to select relevant features from the raw gait cycle data [14]. This process can be time-consuming, prone to human error, and may overlook subtle but crucial patterns in the data. Moreover, classical machine learning algorithms such as SVM, DT, or RF may struggle to capture complex nonlinear relationships inherent in gait cycle data. Parkinson's disease diagnosis often involves subtle changes in gait dynamics that might not be adequately modeled by traditional algorithms, leading to suboptimal performance. Last but not least, classical machine learning models may not generalize well to unseen data or new patients due to their limited capacity to learn intricate patterns and variations present in gait cycle datasets [15]. This lack of generalization could hinder the model's reliability and effectiveness in real-world clinical settings where variability in patient data is common. That is why many deep learning-based models have been presented to solve this problem.

The proposed model of [16] has used 18 one-dimensional (1D) CNN layers to extract features from 1D signals produced by measuring foot sensors. Finally, in order to concatenate the outputs of the 1D-CNN layers, a fully connected network was used for binary classification. While this research focused on the spatial feature, [17] has proposed a model containing long short-term memory (LSTM) parallel layers to extract temporal features from 1D signals. The system is also capable of autonomously assessing the extent of the illness. The proposed mode of [18] created a dual-channel framework integrating Long Short-Term Memory (LSTM) and Convolutional Neural Network (CNN) architectures to comprehend the spatial and temporal patterns inherent in the gait dataset. In [19], a model including two inputs with CNN, LSTM, and an attention mechanism in each input has been designed to extract features from left and right foot signals separately. Also, [20] has used a dual-branch model for feature extraction from left and right foot signals using two CNN layers and two bi-directional LSTM layers in each branch. Then, the output of the two branches is combined, and finally, classification is done with the help of a bi-directional LSTM layer and fully connected layers. The study of [21] first has taken the average of all the left and right foot signals and then produced an image of the frequency changes of the signal for each sample using a spectrogram. Finally, it performed classification by transfer learning approaches such as various CNN models, including AlexNet, ResNet-50, ResNet-101, and GoogLeNet. The study of [22] emphasized the difference in the temporal patterns of the gait cycle between healthy individuals and PD patients and focused on reducing data dimensions by combining sequences of corresponding sensors. The proposed method utilized several interconnected layers containing parallel LSTMs to extract temporal and spatial patterns for binary classification. Also, the study [23] has used the architecture of [22] with the difference that the normalization layer has been removed and each LSTM, unlike before that had one input, included the absolute value of the subtraction of the output of all the same sensors, which improved the accuracy of the model 2.21%. The summary of previous methods is shown in Table 1.

While deep learning-based models for Parkinson's disease diagnosis have shown promise, they are not without limitations. One common drawback is the complexity of the models, particularly those incorporating numerous layers. These deep architectures may suffer from overfitting, especially when dealing with limited datasets, leading to reduced generalization performance on unseen data. Additionally, models like the Network LSTM, which focuses primarily on temporal features extracted through LSTM layers, may overlook crucial spatial patterns inherent in gait data. Conversely, approaches that integrate LSTM and CNN architectures attempt to capture both spatial and temporal features but can introduce computational overhead and increase training time due to having millions of parameters.

Furthermore, reliance on transfer learning approaches with pre-trained CNN models may limit adaptation to specific features of Parkinson's disease gait patterns. Last but not least, previous studies haven't paid attention to the complex dependency between sensors in the feature extraction process. In other words, the amount of pressure applied to each sensor in the walking cycle directly affects the pressure applied to other sensors in its neighborhood. Paying attention to this note can facilitate the process of embedding and analyzing the gait cycle patterns of healthy and diseased individuals.

**Table 1:** Summary of PD detection studies based on gait cycle dataset

| Reference | Classifier | Accuracy (%) |
|-----------|------------|--------------|
| [4] | SVM, RF, DT, K-NN | 91.6, 89.4, 87.21 and 85.1 |
| [11] | K-NN, RF, NB, SVM K-means, GMM | 87.10, 87.10, 83.87,90.32,57.42, and 65.16 |
| [12] | LWRF | 96 |
| [13] | Type-2 neuro-fuzzy system | 97.61 |
| [16] | 1D CNN | 98.7 |
| [17] | LSTM network | 98.6 |
| [18] | CNN-LSTM | 98.06 |
| [19] | DCALSTM | 99.04 |
| [20] | CNN-bidirectional LSTM | 99.22 |
| [21] | AlexNet, ResNet-50, ResNet-101, and GoogLeNet | 93.6, 95.92, 96.07, 94.25 |
| [23] | Multiple layers of parallel LSTMs | 99.87 |

## 3. Data description

In this study, we have used gait cycle signals. This dataset includes Ga [24], Ju [25], and Si [26] from three different research groups that PhysioNet collected. According to Fig. 1, the gait cycle encompasses both the stance and swing phases. Under typical, healthy conditions, the stance phase constitutes approximately 60% of the gait cycle [27].

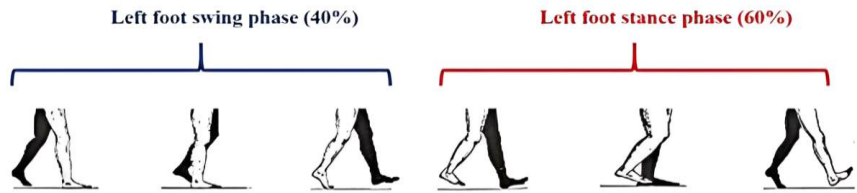

**Fig. 1.** Gait cycle

This dataset comprises gait measurements collected from a cohort of 93 individuals diagnosed with idiopathic Parkinson's disease (PD), with an average age of 66.3 years and 63% male. Additionally, it includes data from 73 healthy individuals serving as controls, with a similar mean age of 66.3 years and 55% male representation. According to Fig. 2, 8 sensors are placed on the soles of each person's feet in different positions to measure force in Newton.

Then, subjects walk at a desired speed for approximately 2 minutes on a smooth path, and during this time, the sensors take samples of people's walking with a sampling frequency of 100. The database also specified the PD severity level of the samples based on the H&Y standard [28] in four levels, including:

- Label of 0: No functional disability (NFD)
- Label of 2: PD without impairment of balance (WOIB)
- Label of 2.5: PD with impairment of balance (WIB)
- Label of 3: PD with impaired postural reflexes (WIPR)

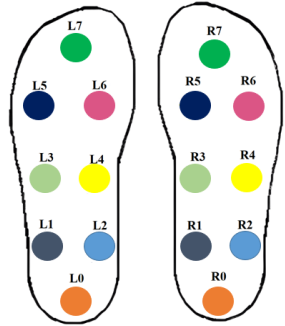

**Fig. 2.** Arrangement of various sensors beneath the feet.

**Table 2:** Details three different gait cycle signal datasets (CO: Healthy Control Subject).

| Reference | Subjects | Male | Female | Total subjects |
|---|---|---|---|---|
| Ga [24] | CO | 10 | 8 | 18 |
| | PD | 20 | 9 | 29 |
| Ju [25] | CO | 12 | 14 | 26 |
| | PD | 16 | 13 | 29 |
| Si [26] | CO | 18 | 11 | 29 |
| | PD | 22 | 13 | 35 |

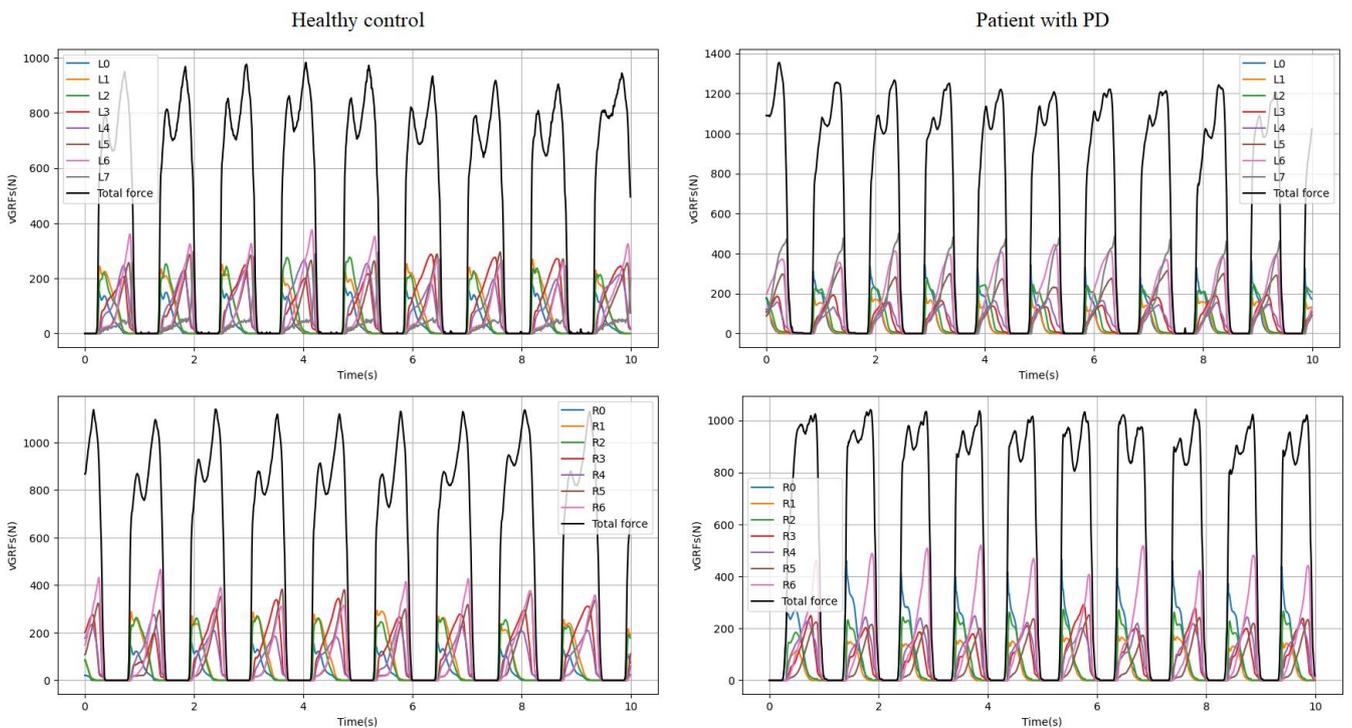

**Fig. 3.** VGRF of left and right feet sensors for the healthy controls and PD patients

The details of the dataset can be seen in Table 2. The dataset contains 19 columns for each person, including time stamp records, signals from 16 sensors, and two columns of total pressure on the left and right soles. Fig. 3 illustrates recorded sequences of data output from diverse sensors of a healthy control participant and an individual with the condition.

## 3.1. Preprocessing

Preprocessing is very important in deep learning-based problems because it is a platform for the effectiveness and efficiency of the model. In this study, in order to improve and facilitate the learning process, we first normalize the input data and reduce its dimensions. Then, we increase the number of samples to reduce the probability of overfitting. Finally, we convert each sample into a graph structure to consider the dependence and complex features between entities (sensors).

### 3.1.1. Normalization and Dimension Reduction

Normalizing data between 0 and 1 offers several advantages that contribute to improved model performance and stability. Firstly, this range ensures that all features are on a consistent scale, preventing certain features with larger magnitudes from dominating the learning process. Normalization facilitates fair comparisons and equitable contributions from each feature during model training by bringing all features within the same numerical range. Additionally, scaling to a bounded interval such as [0, 1] aids in stabilizing the optimization process, as it constrains the magnitude of parameter updates during gradient-based optimization algorithms, thereby fostering smoother convergence and mitigating the risk of overshooting. Therefore, we have used Eq. (1), the Min-Max normalization method, to normalize the values recorded by each sensor.

$$v' = \frac{v - minS}{maxS - minS} \qquad (1)$$

Where:

- $v'$ is new value
- $v$ is the old value
- $minS$ and $maxS$ are the minimum and maximum values of each sensor data.

As Fig. 3 illustrates, there are a great number of zero values in the gait cycle of subjects because of the swing phases. Hence, we calculate the absolute value of the subtraction between the output of each sensor and its counterpart in the other foot. This helps us not only remove zero values but also reduce the information and dimensionality of the data. Therefore, each subject has just eight sequences.

### 3.1.2. Data augmentation

According to Table 2, we have limited samples for model training and testing. It is obvious that there is a clear recognizable relationship between the number of samples and the probability of overfitting. The less the probability of overfitting is, the more generalizable the model is. Therefore, data augmentation plays a crucial role in increasing the performance and robustness of the model.

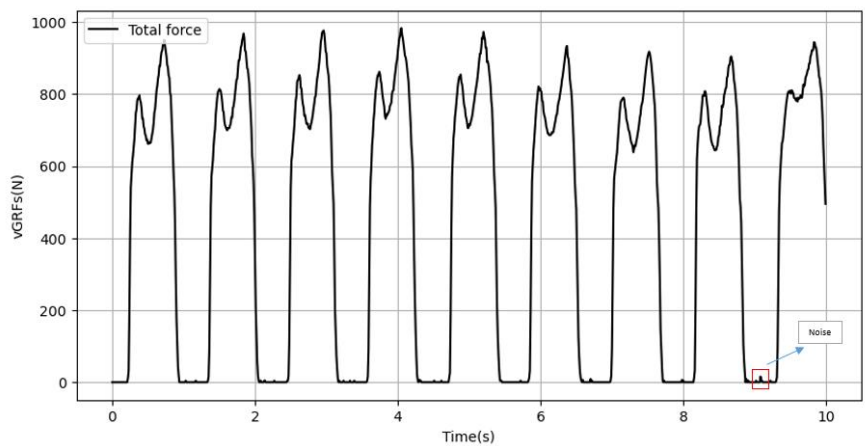

**Fig. 4.** Diagram of the total pressure exerted on the sole during gait cycles

According to Fig. 4, the waveform of the pressure applied to the soles of the feet in healthy and PD samples is an almost repetitive cycle. One cycle per foot starts when the sole hits the ground, and the pressure gradually increases until the sole touches the ground. Then, the pressure decreases until it becomes zero (the sole has no contact with the ground). According to the investigations, each cycle lasts approximately 1.6 seconds, which is equivalent to 160 rows in the data file for each subject. Therefore, to increase the number of samples and remove noise, we have considered every 160 rows of the data file of individuals as a sample. We have not considered the incomplete cycles (less than 160 rows) at the end of the data files. Table 3 shows the number of samples produced in each dataset.

**Table 3:** The segmentation of data samples based on gait cycle

| Dataset | Subjects | Number | Total |
|---|---|---|---|
| **Ga [24]** | CO | 2801 | 8426 |
| | PD | 5625 | |
| **Ju [25]** | CO | 1295 | 7323 |
| | PD | 6028 | |
| **Si [26]** | CO | 2175 | 4800 |
| | PD | 2625 | |

### 3.1.3. Data transformation

As seen in Fig. 2, eight sensors are in different positions on the soles of each person's feet. Pressure changes applied to each sensor affect the output of other sensors located in its neighborhood. As shown in Fig. 5, we have created a graph structure for each gait cycle (160 rows in the data file) to consider the complex dependencies and relationships between nodes (sensors) in the embedding process and also facilitate the analysis of the subject's gait cycle. In this structure, there are undirected and without-weight edges between each node and its neighboring nodes. In other words, edges show the neighborhood relationship between nodes. Also, according to the explanation in section 3.1.1, the data of each node is equal to the absolute value of the subtraction between the outputs of each sensor with the sensor of the same name on the other foot.

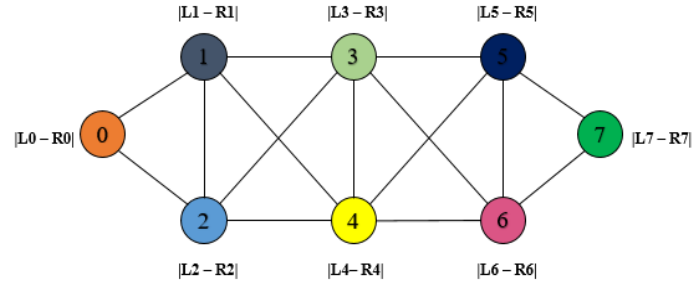

**Fig. 5.** Graph structure of each gait cycle

## 4. Proposed method

### 4.1. Problem statement

In the landscape of Parkinson's disease (PD), the significance of early diagnosis transcends mere clinical expediency, heralding a paradigm shift in the trajectory of patient care and therapeutic outcomes. One of the most important aspects of early diagnosis is the potential to mitigate disease progression and optimize treatment efficacy. Hence, various machine learning methods, especially models based on gait cycle signals, have been presented. One of the most important limitations of the proposed methods is not paying attention to the complex dependencies between sensors of the soles in the embedding process, which leads to a decrease in the model's performance. This study aims to train a model based on deep learning using graph structure samples representing the neighborhood relationship between sensors and step cycle signals.

If $\mathbb{G}$ is a dynamic graph and its snapshots contain $\{G_1, G_2, ..., G_T\}$, then the problem is graph-level binary classification showing Co or PD of the input. To formalize this prediction task, Eq. (2) introduces the inputs and output of the problem, where f is a function that takes in the sequence $\{G_1, G_2, ..., G_{160}\}$.

The structure of the input is the type of static graph-temporal signal. In other words, the arrangement of the network remains constant throughout time (during the gait cycle), but the attributes of the network nodes over time.

$$f(\{G_1, G_2, \ldots, G_{160}\}) = 0 \text{ or } 1 \qquad (2)$$

Tackling the problem using our proposed method not only simplifies the comprehension of complex data patterns by minimizing model parameters but also boosts the embedding process. Additionally, it provides a more detailed examination of the walking behaviors of both healthy and sick individuals by identifying the key sensors involved in the classification process.

### 4.2. Technical background

#### 4.2.1. 1D CNN

CNNs have revolutionized various fields, including computer vision, natural language processing, and signal processing. While commonly associated with image data, CNNs are versatile and extendable to various forms of sequential data, such as time series or one-dimensional signals. 1D CNNs are particularly effective in extracting patterns from sequential data while preserving the local dependencies [29]. The fundamental operation in a 1D CNN is convolution. Convolution involves sliding a filter/kernel over the input signal, multiplying the filter's weights with the corresponding input values, and summing the results to produce a feature map. This process captures local patterns or features within the input signal. Eq. (3) the convolution operation at each step can be expressed as follows:

$$y[i] = \sum_{K=0}^{K-1} x[i + K] \times w[K] + b \qquad (3)$$

Where:

- y[i] is the output at position i.
- x[i + K] represents the input signal at position i + K.
- w[K] denotes the weights of the filter at position K.
- K is the size of the filter/Kernel.
- b is the bias term.

This equation demonstrates how the output at each position is obtained by taking the dot product between the filter weights and the corresponding section of the input signal, followed by adding a bias term.

#### 4.2.2. GRU cell

GRUs are a variant of recurrent neural networks (RNNs) designed to address the vanishing gradient problem and capture long-term dependencies in sequential data more effectively [30]. GRUs achieve this by incorporating gating mechanisms that control the flow of information through the network. The key components of a GRU include an update gate and a reset gate. These gates determine how much information from the previous time step should be passed to the current time step. Mathematically, the computations within a GRU cell can be represented as follows:

$$z_t = \sigma\left(W_z \cdot [h_{t-1}, x_t] + b_z\right) \qquad (4)$$
$$r_t = \sigma\left(W_r \cdot [h_{t-1}, x_t] + b_r\right) \qquad (5)$$
$$\bar{h}_t = \tanh\left(W \cdot [r_t \odot h_{t-1}, x_t] + b\right) \qquad (6)$$
$$h_t = (1 - z_t) \odot h_{t-1} + z_t \odot \bar{h}_t \qquad (7)$$

Where:

- $h_t$ is the hidden state at time step $t$.
- $x_t$ is the input at time step $t$.
- $z_t$ is the update gate that determines how much of the previous state should be retained.
- $r_t$ is the reset gate that controls how much of the previous state should be forgotten.

- $\tilde{h}_t$ is the candidate hidden state.
- $\sigma$ represents the sigmoid activation function.
- $\odot$ denotes element-wise multiplication.
- $W_z$, $W_r$, $W$ are weight matrices, and $b_z$, $b_z$, $b$ are bias vectors.

The update gate $z_t$ decides how much information from the previous hidden state $h_{t-1}$ should be passed to the current hidden state $h_t$, while the reset gate $r_t$ determines how much of the previous state should be forgotten. The candidate hidden state $\tilde{h}_t$ is computed based on the input and the reset gate. Finally, the current hidden state $h_t$ is a linear interpolation between the previous hidden state $h_{t-1}$ and the candidate hidden state $\tilde{h}_t$, controlled by the update gate $z_t$. GRUs are particularly effective in tasks involving sequential data processing, such as natural language processing, time-series prediction, and speech recognition, due to their ability to capture long-range dependencies while mitigating the vanishing gradient problem commonly encountered in traditional RNNs. Fig. 6 illustrates the architecture of a GRU cell.

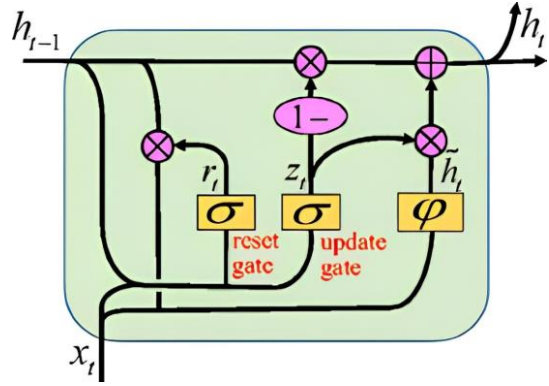

**Fig. 6.** Gait recurrent unit

*4.2.3. GNN*

GNNs have become instrumental in analyzing and understanding graph-structured data. One of the key components of GNNs is the message-passing mechanism, which allows nodes in a graph to exchange information with their neighbors, enabling effective representation learning [31]. The Graph Attention Convolutional (GATConv) layer is a variant of the standard graph convolutional layer that selectively incorporates attention mechanisms to aggregate information from neighboring nodes [32]. This attention mechanism enables GATConv to assign different importance weights to different neighbors, allowing the model to focus on more relevant nodes during message passing. Eq. (8) shows the computation in the GATConv layer can be expressed as follows:

$$h_i^{(l+1)} = \sigma \left( \sum_{j \in N(i)} \alpha_{ij}^{(l)} \cdot W^{(l)} h_j^{(l)} \right) \qquad (8)$$

Where:

- $h_i^{(l)}$ and $h_j^{(l)}$ represent the feature representations of nodes I and j at layer $l$, respectively.
- $N(i)$ denotes the set of neighboring nodes of node $i$.
- $\alpha_{ij}^{(l)}$ represents the attention coefficient between nodes $i$ and j at layer $l$, computed as: $\alpha_{ij}^{(l)} = \frac{\exp(e_{ij}^{(l)})}{\sum_{k \in N(i)} \exp(e_{ik}^{(l)})}$ where $e_{ij}^{(l)}$ is the attention score between nodes $i$ and $j$, computed using a learnable attention mechanism.
- $W^{(l)}$ denotes the learnable weight matrix at layer $l$.

The GATConv layer computes a weighted sum of the neighboring node representations, where the attention coefficients determine the weights. These attention coefficients are computed based on the similarity between node features, allowing the model to focus on more informative neighbors during message passing. By incorporating attention mechanisms, the GATConv layer enhances the expressive power of GNNs, enabling them to capture complex relational patterns within graph-structured data. This makes GATConv and similar attention-based layers valuable tools for various graph-related tasks, including node classification, link prediction, and graph classification.

### 4.3. CGG method

According to Fig. 7, our proposed architecture is tailored for graph-level binary classification tasks, where the input comprises graphs with eight nodes, and each node is associated with signal data of length 160. The architecture is designed to effectively capture both local and global dependencies within the graph structure to make accurate predictions at the graph level. Initially, the model employs a hierarchical feature extraction approach. The first layer utilizes three 1D CNN layers to extract spatial features from the node signals. This step allows the model to capture local patterns within each node's data efficiently. Subsequently, the extracted features are further refined using two GRU layers. GRUs are well-suited for capturing sequential dependencies within the node data, enabling the model to encode temporal information effectively.

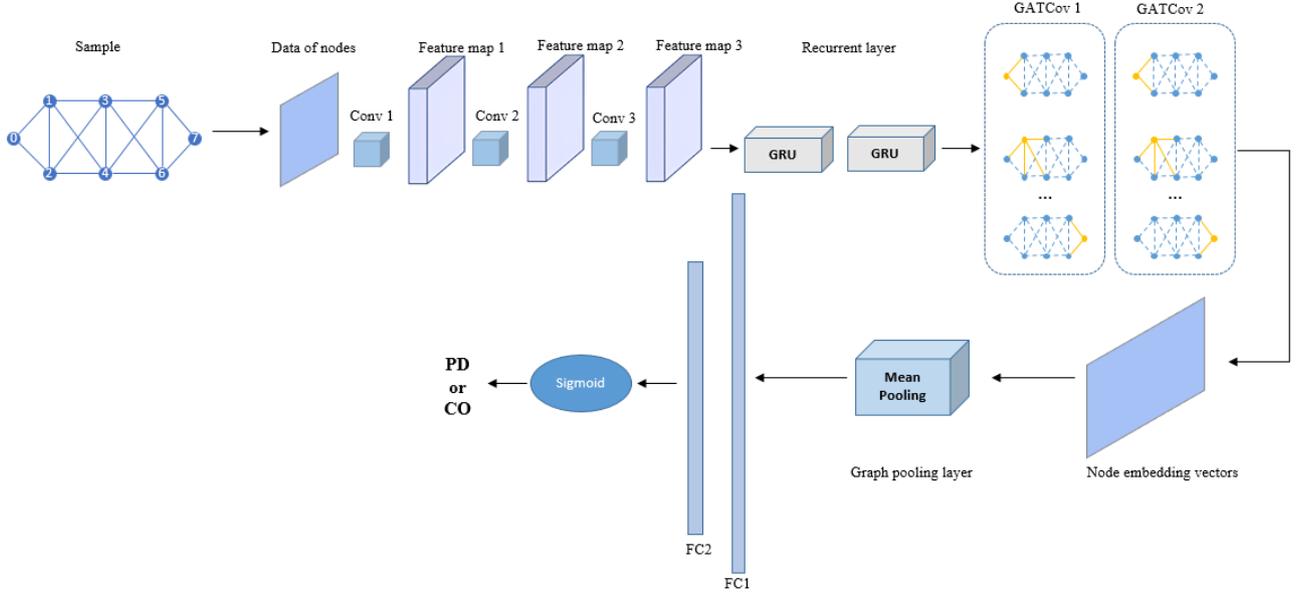

**Fig. 7.** Architecture of the proposed method (CGG)

Following the feature extraction stage, the architecture incorporates two layers of GATCov. GATCov layers facilitate information aggregation from neighboring nodes while considering the covariance between node representations. This step enhances the model's ability to capture relational information within the graph structure. Next, a mean pooling layer is employed to represent the entire graph comprehensively. This layer aggregates the node embeddings obtained from the previous stage, producing a fixed-size vector that encapsulates the collective information from all nodes in the graph. Finally, a fully connected layer is utilized for graph-level classification. This layer processes the aggregated graph representation generated by the mean pooling layer. It produces the output logits, which are subsequently passed through a sigmoid activation function to obtain the final binary classification prediction. It should be noted that a Dropout layer has been used between the proposed layers with a probability of 0.2. We have also used "binary cross-entropy" loss function [33] to train the model defined as Eq. (9) where N is the number of training samples, y is a true label, and $\hat{y}$ is the predicted probability of the positive class. Also, "Adam" algorithm [34] has been used to learn the network's parameters. Specific parameters are shown in Table 4.

$$BCE = -\frac{1}{N}\sum_{k=1}^{N} y_k \log \hat{y}_k + (1 - y_k) \log(1 - \hat{y}_k) \qquad (9)$$

Overall, the proposed architecture integrates multiple layers of neural network modules, including 1D CNNs, GRUs, GATCov, mean pooling, and fully connected layers, to effectively capture both local and global dependencies within the input graphs. Through hierarchical feature extraction, node embedding, and graph-level aggregation, the model demonstrates strong capabilities in performing graph-level binary classification tasks. We have used PyTorch [35] and PyTorch Geometric [36] libraries to implement the proposed model.

**Table 4.** Parameters of the CGG method

| Parameter | Value |
|---|---|
| Graph data shape | $8 \times 160$ |
| Kernel size of Conv1 | $1 \times 3$ |
| Kernel size of Conv2 | $1 \times 3$ |
| Kernel size of Conv3 | $1 \times 3$ |
| Feature map number F1 | 158 |
| Feature map number F2 | 156 |
| Feature map number F3 | 154 |
| GRU output shape | $128 \times 8 \times 256$ |
| GATCov output shape | $128 \times 8 \times 256$ |
| Number of FC1 neurons | 256 |
| Batchsize | 128 |
| Epoch | 140 |
| Learning rate | 0.001 |
| Time step | 160 |

## 5. Experiments

### 5.1. Evaluation metrics

In assessing the performance of classification models, it is imperative to employ a range of evaluation metrics to understand their effectiveness comprehensively. The accuracy, precision, recall, and F1 score metrics offer unique insights into different aspects of the model's predictive capabilities. According to Eq. (10), accuracy is useful for assessing the model's performance. However, it may not be suitable for imbalanced datasets, where the distribution of classes is skewed, as it can be misleading in such cases.

$$Accuracy = \frac{Number\ of\ correct\ Predictions}{Total\ Number\ of\ Prediction} \tag{10}$$

Precision, Eq. (11), measures the proportion of true positive predictions out of all positive predictions made by the model. It highlights the model's ability to correctly identify relevant instances from the total predicted positive instances. Precision is particularly useful in scenarios emphasizing minimizing false positives, such as in medical diagnosis or fraud detection.

$$Precision = \frac{True\ Positives}{True\ Positives + False\ Positives} \tag{11}$$

Recall, Eq. (12), quantifies the model's ability to capture all relevant instances of a particular class within the dataset. It represents the ratio of true positive predictions to the total number of actual positive instances in the dataset. Recall is crucial when missing positive instances, such as identifying critical medical conditions, can have significant consequences.

$$Recall = \frac{True\ Positives}{True\ Positives + False\ Negatives} \tag{12}$$

The F1 score, Eq. (13), is the harmonic mean of precision and recall, providing a balanced measure that considers false positives and false negatives. The F1 score comprehensively assesses the model's performance across different classes by incorporating precision and recall into a single metric. It is particularly valuable in scenarios where achieving a balance between precision and recall is essential, as it penalizes models that favor one metric over the other.

$$F1\ score = 2 \times \frac{Precision \times Recall}{Precision + Recall} \tag{13}$$

## 5.2. Experimental results and discussion

Among the merged data (20549 samples), we have randomly chosen 70% as training data (14384 samples), 15% as test data (3083 samples), and 15% as evaluation data (3082 samples). In Fig. 8, we present a detailed visualization of the training process through a line graph showcasing the fluctuation of accuracy and error metrics over successive epochs for training and validation data, offering insights into the model's convergence and generalization capabilities.

As an overall trend, it is obvious that the training and validation accuracy rate increased over the iterations while the error rate for both decreased. At the beginning of the period, the model's training and validation data accuracy was around 0.69. Within 26 iterations, these figures have risen gradually to achieve 0.96. After three epochs, both of them experienced a sharp fall when they reached 0.78. Then until the epoch of 57, they have climbed moderately to achieve 0.98. Finally, after improving the model's accuracy for the training data compared to the evaluation data, the model's accuracy for both of them has finished at just under 1 after slowly changing. On the other hand, the validation loss has begun to progressively decrease from a value of 0.4 to surpass the training loss at epoch 57, after a sudden increase to a value of 0.37 at epoch 26. Then, these figures experienced a lot of fluctuation without any increase or decrease until finished at well over 0.

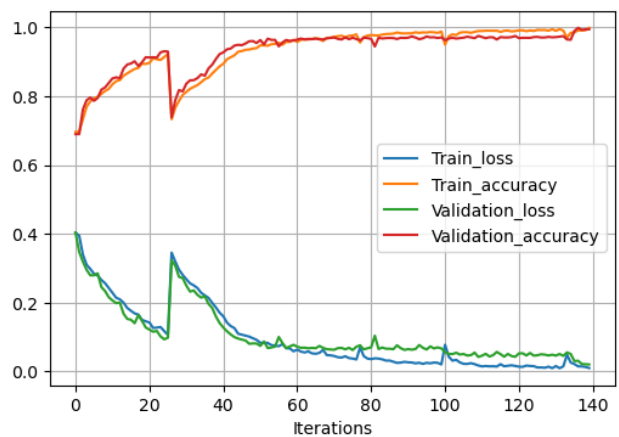

**Fig. 8.** The changing of accuracy and loss values during the training process for training and validation datasets

The confusion matrix, Fig. 9, provides a detailed breakdown of the performance of our classification model, illustrating the correspondence between predicted and actual class labels for the test dataset. In this matrix, the rows represent the actual classes, while the columns represent the predicted classes. The first row pertains to class 0 (CO), and the second row corresponds to class 1 (PD).

Our model demonstrates strong performance in correctly identifying both healthy individuals and patients, as evidenced by the high numbers of true positives and true negatives. However, it exhibits a slight tendency towards false positives, with 9 instances incorrectly classified as patients when they were actually healthy. Conversely, there were 6 instances where patients were misclassified as healthy individuals, indicating a relatively minor misclassification rate for this class.

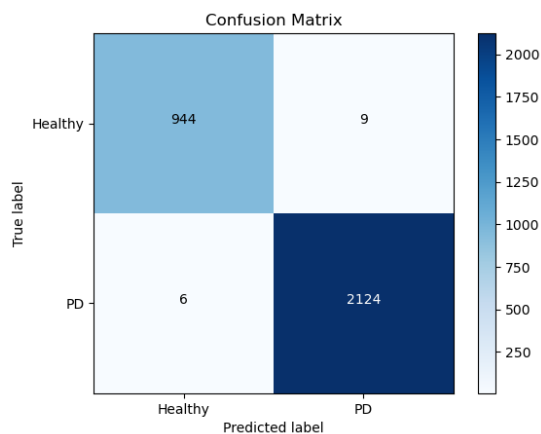

**Fig. 9.** Confusion matrix of the proposed model

In the Receiver Operating Characteristic (ROC) diagram presented in Fig. 10, the area under the curve (AUC) is observed to be 1, indicating exceptional discriminatory power and robust performance of the classification model. The ROC curve elegantly illustrates the trade-off between true positive rate (sensitivity) and false positive rate (1 - specificity), showcasing the model's ability to differentiate between classes across various classification thresholds.

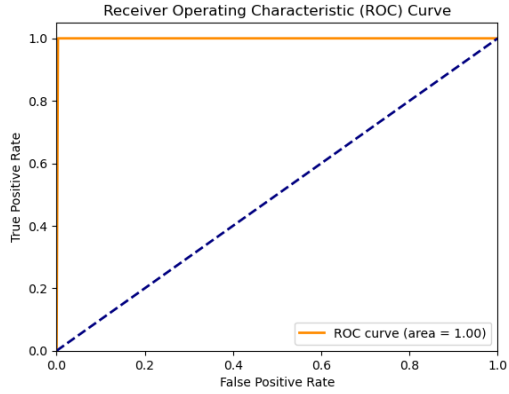

**Fig. 10.** ROC curve of the proposed model

To evaluate the efficacy of our binary classification model, we conducted a comparative analysis against several baseline methods commonly used in similar tasks. The performance metrics of each model were assessed across various evaluation criteria, including accuracy, precision, recall, F1 score, and the number of parameters. Table 5 provides a comprehensive overview of the comparative results, highlighting the strengths and weaknesses of our model in relation to the established baselines.

The findings underscore the superior performance of SVM and LWRF algorithms compared to traditional machine learning approaches. Nevertheless, the reliance on non-optimal and statistical methods for extracting feature vectors in classical models has limited their effectiveness, with most exhibiting subpar performance less than 90%.

**Table 5.** Comparison of the proposed model performance with other baselines based on all datasets.

| Classical machine learning algorithms | Accuracy (%) | Precision (%) | Recall (%) | F1 score (%) | Number of parameters |
|---|---|---|---|---|---|
| SVM | 96.60 | 94.10 | 93.10 | 93.59 | - |
| LWRF | 96.89 | 96.55 | 97.02 | 96.78 | - |
| DT | 85.23 | 83.97 | 86.05 | 84.99 | - |
| K-NN | 89.01 | 89.08 | 90.05 | 89.56 | - |
| K-means | 57.42 | 58.52 | 57.88 | 58.20 | - |
| NB | 83.87 | 85.31 | 83.54 | 84.42 | - |
| GMM | 65.16 | 65.94 | 64.73 | 65.33 | - |
| Neuro-fuzzy system | 97.61 | 97.58 | 99.02 | 98.30 | - |
| **Methods based on deep learning** | | | | | |
| 1D CNN | 98.70 | 100 | 98.1 | 99.04 | 857120 |
| LSTM network | 98.60 | 99.10 | 98.23 | 98.95 | 1119529 |
| CNN-LSTM | 98.60 | 98.06 | 98.81 | 98.43 | 66505078 |
| DCALSTM | 99.04 | 98.09 | 99.03 | 98.55 | 89201292 |
| CNN-bidirectional LSTM | 99.22 | 98.04 | 100 | 99.01 | 2022785553 |
| ResNet-101 | 96.07 | 96.52 | 95.28 | 95.89 | 44500000 |
| Multiple layers of parallel LSTMs [23] | **99.87** | - | - | 96.66 | < 857120 |
| The proposed model (CGG) | 99.51 | 99.57 | **99.71** | **99.64** | 909837 |

In contrast, deep learning methodologies automate feature extraction from input data and consistently outperform classical methods. The proposed method considers spatial and temporal features and intricately captures complex sensor interactions within the gait cycle. Although the proposed model's accuracy is less than [23], our model's F1 score is higher. Since the lower F1 score and higher accuracy show that the model has a higher false positive or false negative rate, it can be said that the proposed model is better than [23] for two main reasons.

On the one hand, accuracy can be misleading in datasets such as the dataset used in this study, where one class significantly outweighs the others (class imbalance). Precision and recall provide more insight into the model's performance in such cases. On the other hand, in some applications, such as medical diagnoses, certain types of errors have more severe consequences than others. For example, in medical diagnoses, a false negative (missing a disease when it's present) can be more harmful than a false positive (diagnosing a disease when it's not present). Therefore, maximizing recall to reduce false negatives might be more important than overall accuracy. Overall, the proposed model not only performs significantly better in the binary classification of PD and CO subjects compared to alternative approaches but also has fewer parameters than models that use CNN and LSTM to extract spatial and temporal features.

According to Fig. 7, in the architecture of the proposed model, the GATCov2 layer allocates a 256-dimensional vector for each node based on the importance coefficient. To identify the influential sensors in the classification process and to facilitate the analysis of the step cycle of different groups, we have converted the vector of each node into a color code. The warmer a node's color is, the more important it is. Although no consistent pattern was observed in the color of nodes for different samples, according to Fig. 11, there is a definite link between the severity of the PD and the sensors in the heel area, nodes 0, 1, and 2. The higher the severity of PD is, the more patients use their heels to maintain balance during the gait cycle. This is even though sensors in the toe area, nodes 5, 6, and 7, usually have a greater effect on identifying healthy subjects. Also, nodes 3 and 4 have usually less impact on classifying subjects.

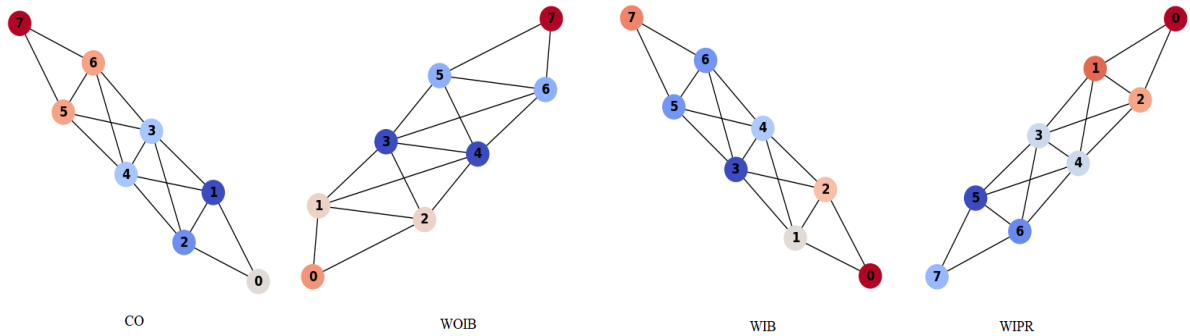

**Fig. 11.** Influential nodes of different classes in the binary classification process

## 6. Conclusion

Parkinson's disease is one of the most common neurological disorders in the world. Early diagnosis of this disease can greatly reduce treatment costs and make its management easier. Among these, several methods based on machine learning have been presented for early diagnosis, which uses different methods such as gait cycle signals, audio signals, brain imaging, etc. Gait cycle signals are a more appropriate option for evaluating individuals due to their spatial-temporal features and lower cost in diagnosis rather than other ways. However, the previous methods face some problems, for example, the probability of overfitting because of the low number of samples in the gait cycle dataset, increasing the training time due to having millions of parameters, and not paying attention to complex dependencies like the effect of each sensor on its neighbor's counterpart in the embedding process.

In this study, we have considered every 160 rows, a gait cycle, of the data files as a sample to increase the number of samples. Then, we changed the data structure to a graph to consider the complex relationships between different sole sensors. Therefore, in the architecture of the proposed method, we have used 1D-CNN, GRU, and GNN layers to pay attention to the spatial-temporal features and complex dependencies between nodes. The results show that the proposed method has performed better than other baselines in binary classification. The proposed model's accuracy, precision, recall, and F1 score are equal to 99.51, 99.57, 99.71, and 99.64, respectively.

In light of the findings presented in this study, there are several promising avenues for future research. As the future trends of this study, we propose 1- to study vocal impairment by converting each sample as a graph, 2- to quantify the severity of Parkinson's disease, and 3- to investigate the precision of this method to diagnose patients with similar diseases like Huntington's disease.

## Credit authorship contribution statement

**Alireza Rashnu:** Methodology, Software, Conceptualization, Validation, Formal analysis, Theoretical analysis, Investigation, Data curation, Visualization, Writing - original draft, Writing – review & editing. **Armin Salimi-badr:** Methodology, Conceptualization, Supervision, Writing – review & editing, Project administration, Theoretical analysis.

**Funding**  The completion of this article is solely attributed to the collaborative efforts and dedication of the authors involved. We are grateful for the collaborative spirit and academic environment that has enabled us to undertake and complete this study.

## Compliance with Ethical Standards